# Freely controllable single-optical-frequency comb for highly sensitive cavity ring-down spectroscopy

NORIHIKO NISHIZAWA,[1,*] SHOTARO KITAJIMA,[1] NINGWU LIU,[2] RYOHEI TERABAYASHI,[2] DAIKI HASHIMOTO,[2] HISASHI ABE,[3] AND HIDEKI TOMITA[2]

[1]*Department of Electronics, Nagoya University, Furo-cho, Chikusa-ku, Nagoya 464-8603, Japan*
[2]*Department of Applied Energy, Nagoya University, Furo-cho, Chikusa-ku, Nagoya 464-8603, Japan*
[3]*National Metrology Institute of Japan (NMIJ/AIST), AIST Tsukuba Central 3, Tsukuba 305-8563, Japan*
*\*nishizawa.norihiko.w4@f.mail.nagoya-u.ac.jp*

**Abstract:** Direct comb spectroscopy is a useful tool for obtaining highly accurate spectroscopic information. However, as the number of comb modes is very large and the optical energy is dispersed over them, the optical energy per each comb mode is ultrasmall, limiting the sensitivity of highly sensitive spectroscopy. If we can concentrate the optical energy into the comb modes that only overlap with the absorption spectra, we can demonstrate drastic improvements in its measurement sensitivity. In this study, we developed a freely controllable optical frequency comb source based on the spectral peak phenomenon. The comb modes overlapping the $CH_4$ absorption spectra were transformed into background-suppressed spectral peaks at the nonlinear loop mirror using a $CH_4$ gas cell. Coherence-preserving power scaling of the generated comb was demonstrated using a fiber Raman amplifier. Subsequently, only the single-comb mode was filtered using a newly developed spectral filter with an ultrahigh resolution. The maximum optical power of a single comb was estimated to be more than 10 mW. The ring-down decay signal from the high-finesse optical cavity was measured using a single selected mode of the generated controllable comb. As a demonstration, the $2\nu_3$ bands of the $CH_4$ absorption spectra were accurately measured by comb-mode-resolved, cavity ring-down spectroscopy (CRDS) with high sensitivity up to $4.2 \times 10^{-11}$ cm$^{-1}$. This sensitivity is two orders of magnitude higher than that of previously reported comb-based CRDS. The residual was only 0.29 %, indicating the high accuracy of the proposed spectrometer for molecular spectral analysis. This approach can be extended to other wavelength ranges and is useful for highly sensitive, high-resolution, comb-resolved spectroscopy.

## 1. Introduction

Optical frequency combs (OFCs) represent a breakthrough in the field of metrology [1–2]. OFCs have been applied in many fields, such as in optical frequency standards, high-resolution and high-speed spectroscopy, and absolute distance measurements. Spectroscopy is one of the most successful and extensively used OFC applications. In particular, direct frequency comb (DFC) spectroscopy—such as dual-comb spectroscopy, Fourier transform spectroscopy (FTS), and Vernier techniques- directly utilizes the OFC as the light source and has become a useful tool for absorption spectroscopy with comb-mode-resolved resolution [3–5]. Additional sensitivity enhancements of DFC-based spectroscopy may be achieved through the integration of cavity-enhanced techniques.

Cavity ring-down spectroscopy (CRDS) [6] is one of the most commonly used cavity-enhanced techniques for obtaining information and dynamic changes in trace substances. Owing to the development of modern distributed feedback (DFB) lasers and highly reflective mirrors, tunable diode laser-based CRDS can easily achieve exceptionally long optical path lengths spanning more than several kilometers. In addition, intensity fluctuations of the incident laser can be suppressed by measuring the decay rate of the light inside the cavity over several tens of microseconds, without the need for any modulation scheme. These advantages of CRDS have lowered the detection limits for trace species to the nmol/mol (ppb) and (more recently) to the pmol/mol (ppt) mole fraction levels [7–12]. However, DFB lasers have narrow spectral coverage, which limits their ability to detect different gas species with absorption features spanning widely separated wavelength regions when using a single laser. In contrast, integrating a DFC with CRDS can overcome these limitations by taking advantage of the OFC's broad operating bandwidth. This enables the efficient detection of target molecules in complex gas matrices, such as those encountered in breath analysis [13,14], environmental research [15–17], and molecular kinetics research [18–20].

Nevertheless, only a few DFC-based CRDS studies [21–24] have been reported to date. This is likely because the large number of comb modes is not inherently well-suited, particularly for CRDS, necessitating additional mechanisms to select an appropriate subset of modes or to analyze multiple ring-down times simultaneously. For example, Thorpe et al. [21] employed a monochromator for the selection of comb modes. This was the first reported DFC-based CRDS that achieved a sensitivity of $2 \times 10^{-8}$ cm$^{-1}$ with an averaging time of 1 s. The spectral resolution was limited to 25 GHz owing to the resolution of the monochromator; accordingly, this did not resolve each comb mode as the repetition frequency ($f_{rep}$) was 380 MHz. Lisak et al. [22] developed dual-comb CRDS with a spectral resolution of 1 GHz. The ring-down signals for 22 cavity modes were simultaneously acquired by matching the repetition rate of the OFC ($f_{rep}$) to four times the free spectral range (FSR) of the cavity. The resulting interferograms were analyzed using Fourier transform techniques, achieving a sensitivity of $2.6 \times 10^{-9}$ cm$^{-1}$ (normalized to a measurement time of 1 s) across the 22 spectral elements. Chen et al. [23] applied comb modes generated from an interband cascade laser (approximately 40 modes) to CRDS. Multiple ring-down times were sequentially measured within a short time window using a Vernier filtering technique, achieving a sensitivity of $4.3 \times 10^{-8}$ cm$^{-1}$ (normalized to a measurement time of 1 s and one spectral element). A higher spectral resolution of 250 MHz was achieved by Dubroeucq et al. [24] who used a DFC-CRDS system combined with fast-scanning FTS. A sensitivity of $1.1 \times 10^{-9}$ cm$^{-1}$ was demonstrated by averaging 15 successive datasets with a total acquisition time of 1 h and 19 min.

However, in all the aforementioned DFC-CRDS studies, the inherently low power of each narrow comb mode (<100 µW) severely limited the feasibility of employing a cavity with a higher finesse, owing to the very low laser power transmission through the cavity, thereby restricting the achievable sensitivity. Hence, for the practical application of DFC to CRDS, both power scaling and selective comb mode extraction are critical challenges. A major obstacle to efficient power scaling for OFC-based spectroscopy is that a large part of comb modes do not overlap with the absorption lines of the target molecules and thus do not contribute to the spectroscopic signal. Therefore, if we can selectively generate and amplify only the comb modes that coincide with the absorption lines, considerable improvements in power efficiency and signal-to-noise ratio can be expected.

So far, some researchers have investigated the mode selection of OFCs using Fabry–Perot etalons and ring resonators [25-28]. They selected periodic resonant modes with a few GHz spacing using cavity properties. A single-comb mode filtering technique using stimulated Brillouin scattering in optical fibers has been demonstrated by some groups [29-32]. Because it requires a few-kilometer-long fiber and a high-power CW laser, the wavelength range has been limited to within the telecom band, in which an Er-doped fiber amplifier (EDFA) and low-loss single-mode fiber can be used. Because of the need for the precise control of wavelength-tunable LD, this technique has not yet been applied to spectroscopic applications. If we realize a controllable single-comb source while maintaining the excellent properties of the stabilized comb, it can become a useful light source for highly accurate and sensitive spectroscopy.

In 2020, Nishizawa and Yamanaka discovered the novel phenomenon of periodic spectral peaking in optical fibers [33]. Sharp spectral peaks can be generated at the targeted wavelength using a spectral filter or molecular gas, wavelength-adjusted ultrashort pulse, and optical fiber

[33–37], as shown in Fig. 1. This phenomenon enabled not only the selective extraction of comb modes but also efficient power enhancement for the measurement of target molecules.

In this study, we successfully demonstrate the effectiveness of the freely controllable OFC for CRDS. A background-suppressed, intense, and sharp spectral peaked comb was generated at 1.65 μm, customized for the measurement of the target molecule $CH_4$, based on a 206 MHz high-repetition rate fiber laser comb, spectral peaking, and coherence-preserved amplification techniques. The generated spectral peak comb was introduced into a high-resolution spectral filter, and only a single comb mode was selectively filtered out. To the best of our knowledge, this is the first demonstration of DFC-CRDS capable of selecting a single comb mode among a very large number of modes (more than 20,000 modes in this study). The selected single comb was then amplified to a power level exceeding 10 mW and used as a light source for CRDS measurements using an optical cavity finesse of 244,000. As a result, the highest power per comb mode and the highest spectral resolution ever achieved for the DFC-based CRDS were realized, and the highly sensitive absorption spectroscopy of $CH_4$ gas was successfully demonstrated. The characteristics of the developed comb source and the performance of the CRDS system using this source were examined experimentally.

## 2. Experimental

### 2.1 Controllable optical frequency comb source

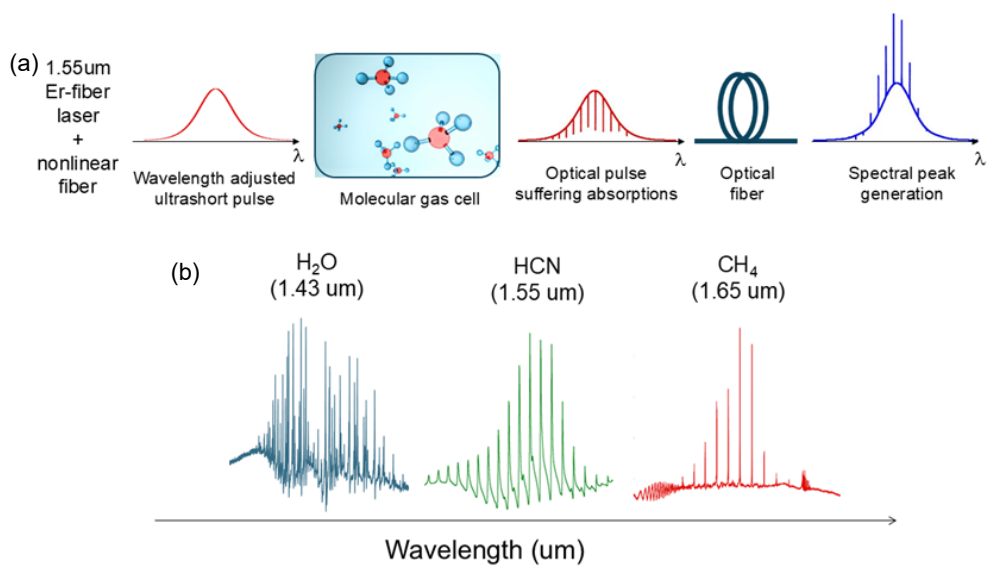

Fig. 1. (a) Principle of spectral peak generation using wavelength-adjusted ultrashort pulses, molecular gases, and optical fibers, and (b) spectra of generated spectral peaks.

Figure 1(a) shows the scheme of spectral peak generation using wavelength-adjusted ultrashort pulses, molecular gases, and optical fibers. When the ultrashort pulses suffering molecular gas absorptions propagate along optical fibers, the absorption dips are converted into spectral peaks using the induced nonlinear phase shift. Figure 1(b) shows the examples of generated spectral peaks using molecular gases. Using the spectral peaking phenomenon with molecular gas, we can filter out only the comb modes that match the absorption spectra of the molecular gas. Based on this phenomenon, we developed a freely controllable comb source for highly sensitive absorption spectroscopy of molecular gases. By adjusting the wavelength to the targeted molecular absorptions, we can generate spectral peaks from the visible to the IR wavelength regions.

Figure 2 shows the configuration of a freely controllable comb source. It consists of a (1) stabilized comb, (2) wavelength shifter, (3) spectral peak generator, (4) coherence-preserving fiber Raman amplifier, and (5) comb mode filter.

To realize a freely controllable narrow-linewidth comb, we started with a stabilized fiber-laser comb. An all-polarization-maintaining fiber-laser-based comb with a repetition frequency of 206 MHz was developed. The oscillation wavelength was 1.55 μm. The details are provided in Ref. [38].

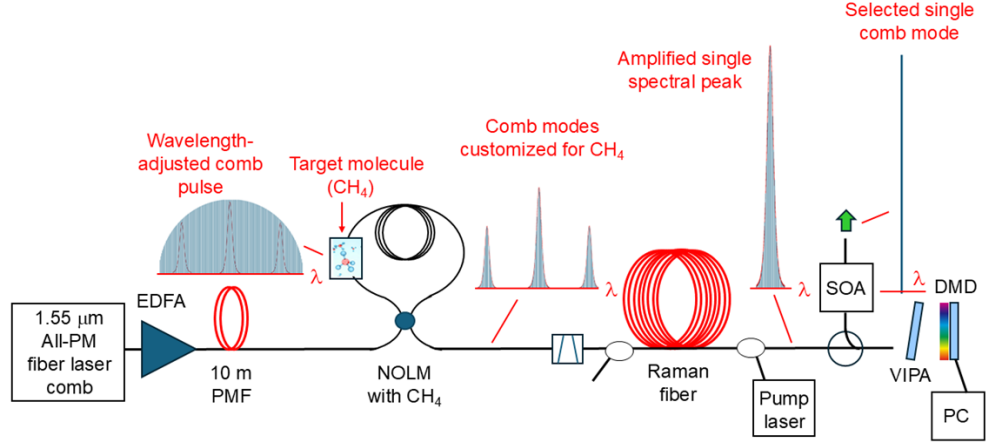

Fig. 2. Configuration and principle of freely controllable optical frequency comb. NOLM, nonlinear loop mirror; SOA, semiconductor optical amplifier.

The output pulse was then amplified with the Er-doped fiber amplifier (EDFA), and a Raman-shifted soliton pulse was generated at 1650 nm, as shown in Fig. 3(a). The pulse width was 100 fs, and the average power was 30 mW. The spectral width was 42 nm, and the corresponding comb mode number was >20,000. The generated soliton pulse was then introduced into a nonlinear loop mirror with a $CH_4$ gas cell [36]. The input pulse was divided at a 50:50 ratio at the fiber coupler. The clockwise pulse passed through the $CH_4$ gas cell at the beginning and then propagated along the 30 m fiber loop, and spectral peaking was induced. The dividing ratio and the loop length were designed by numerical and experimental analyses to obtain intense spectral peaks with high signal-to-background ratio (SBR) [36]. The counterclockwise pulse first propagated along the fiber loop and finally passed through the $CH_4$ gas cell, and the ultrashort pulse with $CH_4$ absorption returned to the coupler. The two counter propagated pulses overlapped at the fiber coupler. The subtracted components emerged from the output port owing to the interference; as a result, background-suppressed intense spectral peaks were obtained, as shown in Fig. 3(b). Multiple spectral peaks with well-suppressed backgrounds are observed stably without any feedback control. The SBR was approximately 30 dB, and the spectral width was as narrow as 32 pm at the center spectral peak. As the spectral peaks were generated from the absorption spectra, the spectral shapes exhibited high long-term stability. Interestingly, an intense spectral peak whose spectral intensity was almost six times larger than that of the original pulse spectra was obtained by soliton spectral compression and spectral peaking. This is a good advantage of the spectral peaking phenomenon.

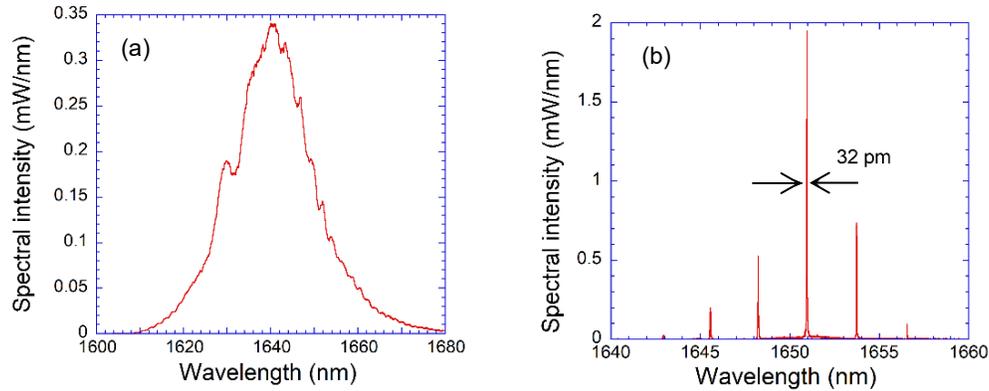

Fig. 3. Optical spectra at the (a) input and (b) output of NOLM equipped with a $CH_4$ gas cell. The gas pressure was 13.3 kPa, and the length of the gas cell was 16.5 cm.

Subsequently, one of the spectral peaks was selected using a wavelength-tunable band-pass filter, passed through an optical isolator, and introduced into a lab-made fiber Raman amplifier for power scaling [37]. Using the fiber Raman amplifier, we can demonstrate power scaling at a broad range spanning from the visible to the IR regions. A Raman fiber (OFS Raman fiber with a length of 1 km) with a small core diameter of 4 μm and normal dispersion properties was used as the gain fiber. This Raman fiber was backward-pumped with a combination of a 1.55 μm multimode LD (Thorlabs SL-1550) and a high-power, double-clad EDFA (Pritel EA-30). A wideband phase modulator (Thorlabs LN65S-FC) was used to broaden the pump spectra and suppress stimulated Brillouin scattering.

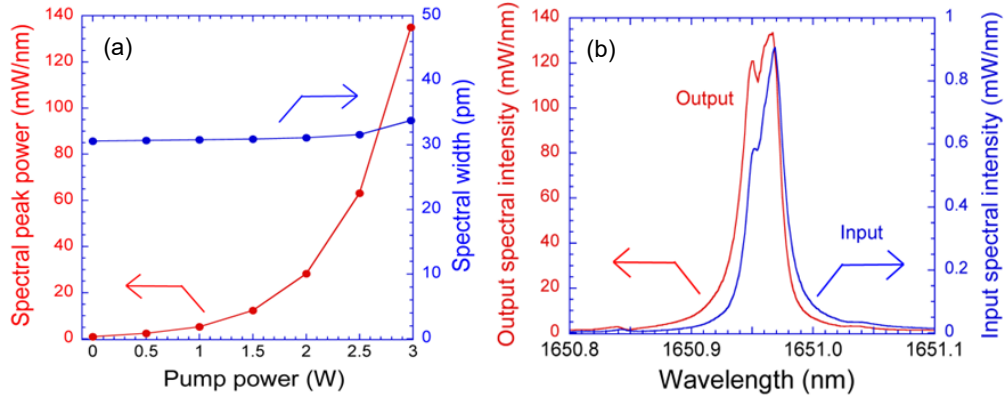

Fig. 4. (a) Variation of peak power and spectral width at fiber Raman amplifier as a function of pump power, and (b) input and output spectra at the fiber Raman amplifier. The spectral intensity of the output pulse is 140 times larger than that of the input pulse.

Figure 4(a) shows the variations in the spectral peak power and spectral width as functions of pump power. As the pump power increases, the spectral peak power increases almost exponentially. The maximum gain was approximately 22 dB (~140), and the highest spectral peak power was 140 mW/nm, which is ~500 times larger than that of pump pulse shown in Fig. 3a. The spectral width was almost constant for the pump power and slightly increased when the pump power increased to value >2.5 W. This spectral broadening is attributed to self-phase modulation (SPM). Figure 4(b) shows the optical spectra before and after amplification. In the previous work, we used a 200 Torr $CH_4$ gas cell in NOLM, and the spectral width of the generated spectral peak was 58 pm at full width at half maximum (FWHM). In this case, owing to SPM, the spectral width broadened to 110 pm and the peak wavelength shifted slightly to the shorter wavelength side by 50 pm under maximum pumping condition [37]. In this work, we used a 100 Torr $CH_4$ gas cell in NOLM, and the spectral width of the generated peak was only 32 pm at FWHM. Consequently, since the temporal width was broader than the previous one, the magnitude of the spectral broadening was reduced to 2 pm, and the wavelength shift was 5 pm. Ideal amplification of the spectral peak was demonstrated. The corresponding comb mode number to the spectral width was ~30.

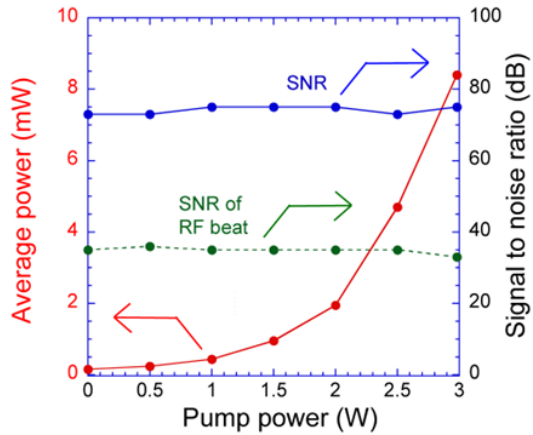

Fig. 5. Variation of signal-to-noise ratios (SNRs) for fundamental frequency and radiofrequency (RF) beat and average power as a function of pump power.

The noise properties of the output spectral peaks were evaluated in the radiofrequency (RF) domain using an RF spectrum analyzer (Anritsu MS2840A). The signal-to-noise ratio (SNR) of the fundamental frequency was observed using a fast-pin photodiode (Thorlabs FPD310-FC-NIR) to evaluate the noise properties of the amplified spectral peaks. The RF beat spectra between the amplified spectral peak and stable CW-LD (Santec SL-510) were also observed to examine coherence properties. The two beams overlapped using a single-mode fiber coupler and were detected using a balanced photodetector (Thorlabs PDB480C-AC).

Figure 5 shows the variations in the average power and SNR of the fundamental frequency and RF beat spectra. The average power increased almost exponentially, similar to the spectral peak. The maximum average power was 8.5 mW. The SNRs were very stable with respect to the pump power. The magnitude of the SNR for the fundamental frequency was 75 dB and that of the RF beat was 35 dB. As a result of SNR measurements, we confirmed that low-noise and coherence-preserving power scaling were achieved in the developed fiber Raman amplifier.

The RIN was measured using an RF spectrum analyzer with the single-sideband measurement method. The phase noise was examined using the RF spectrum analyzer's function. Overall, slight noise increments of ~10 dB were observed in the high-frequency region (>10 kHz) for both the RIN and the phase noise. This magnitude of noise increase is negligible for the CRDS measurement.

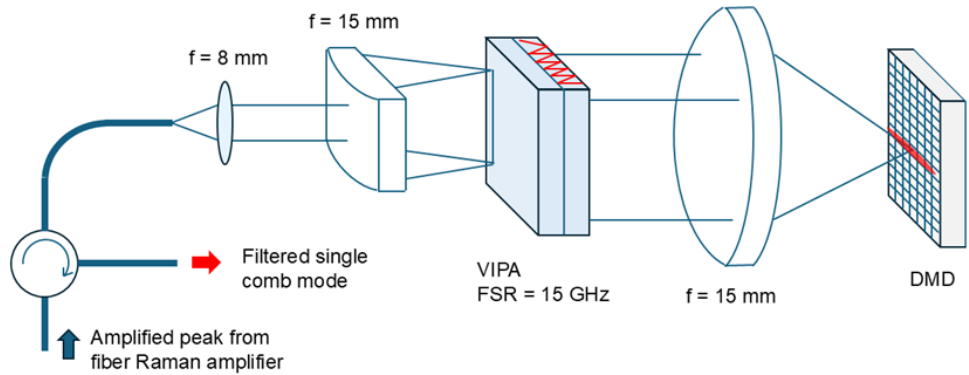

Fig. 6. Configuration of high-resolution spectral filter using virtually imaged phased array (VIPA) and DMD.

Subsequently, a high-resolution spectral filter was developed to realize only one comb-mode filtering scheme. Figure 6 shows the configuration of the proposed spectral filter. It consists of a virtually imaged phased array (VIPA), digital micromirror device (DMD), and optical lenses. The amplified spectral peak was collimated with an asymmetrical lens and coupled to a VIPA filter (LightMachinery OP-6721-6.74-8) using a cylindrical lens (Thorlabs LJ1934L1-C). The FSR of the VIPA was 15 GHz, which corresponds to a spectral range of 138 pm at $\lambda = 1650$ nm. This spectral width covers only one spectral peak in this wavelength range. The input beam was reflected multiple times inside the VIPA filter, and the optical spectra were spatially dispersed at the output port [39]. The output beam was collimated using an asymmetric lens (Thorlabs AC508-150-C-ML) and focused onto the surface of the DMD device (Texas Instruments DL4100). The DMD consisted of micro-mirrors (size of ~10 μm comprising 1280 × 800 pixels), and each mirror was controlled independently via a driver board connected to a personal computer (PC). As written in the Method section, the optical system was designed and carefully optimized to achieve high resolution. A high-spectral resolution of 1.5 pm was achieved with this spectral filter, which corresponds to 165 MHz. Thus, we can independently separate the comb modes with a 206 MHz frequency interval. The reflected beam at the DMD returns to the same optical axis and is coupled back to the single-mode fiber. The filtered beam was separated using an optical circulator (Thorlabs 6015-3-APC) and directed toward the output port. The total optical loss of the filter was approximately 15 dB.

Thus far, VIPA has only been used for spectral mode control and comb mode measurements with spacing greater than 0.8 GHz [40–43]. To the best of our knowledge, this is the first instance of single-comb mode filtering using a VIPA device. The 206 MHz comb mode spacing is also the narrowest and is very useful for spectroscopy. The FSR is one of the limiting factors for realizing single comb mode filtering using a VIPA device. In this work, we used spectral peaking to generate a very narrow, peaked spectral comb whose spectral width was within the FSR of the VIPA device. As a result, we successfully achieved single-comb-mode filtering by combining spectral-peak generation, the VIPA device, and the DMD.

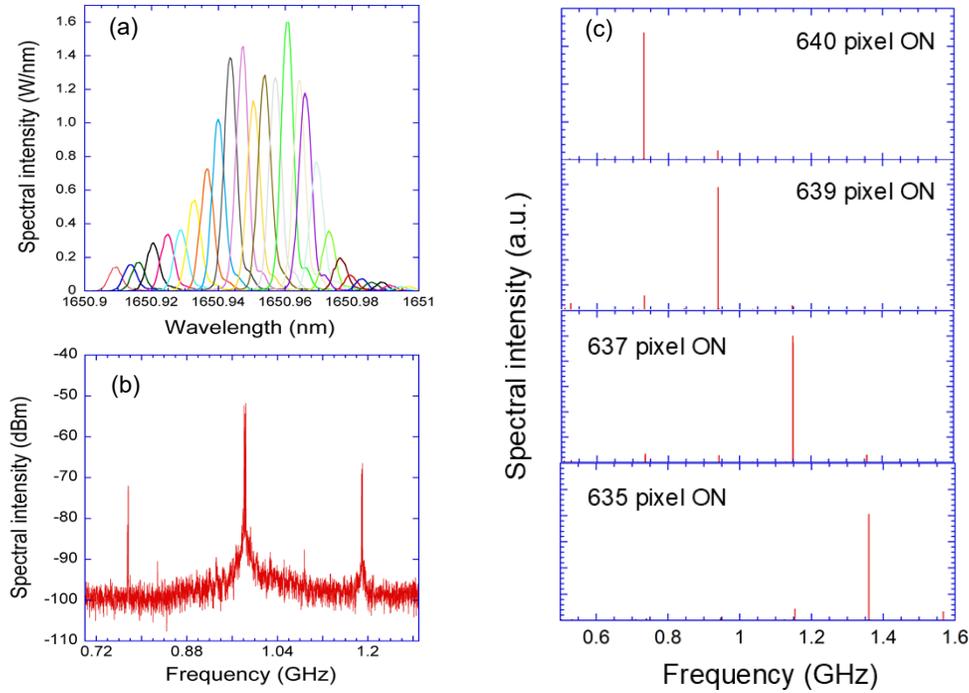

Fig. 7. Variations of output spectra of freely controllable OFC. (a) optical spectra, (b) RF beat spectra on a linear scale, and (c) example of RF beat on a log scale.

The output of the high-resolution filter was introduced into two cascaded semiconductor optical amplifiers (SOAs), and power scaling of the comb was achieved. Figure 7(a) shows the optical spectral variation at the outputs of the SOAs when the ON pixel in the DMD was changed. Herein, only two vertical pixel lines were turned on. Optical spectra were recorded using a high-resolution optical spectrum analyzer. The observed spectral width was slightly broadened by the resolution bandwidth limit, but narrow spectra were filtered out continuously and separately. The highest spectral peak power was 1.6 W/nm, and the corresponding average power was 57 mW. The total net gain of the SOAs was approximately 25 dB, and no spectral degradation was observed in the SOA output. The SNR of the amplified comb was in the range of 60–70 dB. The noise properties were almost preserved during the SOA amplification process. We examined the RIN and phase noise of the fiber Raman amplifier and SOA. Overall, slight noise increments of ~10 dB were observed in the high-frequency region (>10 kHz) for both RIN and phase noise. The effect of this noise increase is negligible for the CRDS measurement.

Figure 7(b) shows one example of the observed RF beat spectra. The main peak was 10 dB higher than that of the side modes. The SNR of the main filtered mode was 35 dB, which was the same as the output of the fiber Raman amplifier. Figure 7(c) shows the variations in the RF spectra of the filtered comb modes on a linear scale. By changing the turned-on pixels in the DMD, we successfully filtered out only one comb mode and changed the comb mode stepwise. Using the driver software on a PC, we can rapidly change the filtered mode. The changing time of the DMD pixel pattern was set to 10 ms, and rapid control of the comb mode was achieved. To the best of our knowledge, this is the first demonstration of switchable single-comb mode filtering based on a stabilized comb source. Several separate modes can be filtered simultaneously as well, which is difficult to achieve when other techniques are used.

In Fig. 7(b), the side-mode suppression ratio was approximately 10 dB, and 33.33–50 % of the optical power was concentrated on the single comb mode. Considering the background noise, the maximum optical power of the single comb was estimated to be >10 mW. To the best of our knowledge, this is the highest power per comb mode for spectroscopic applications.

*2.2 Cavity ring-down spectroscopy using controllable comb source*

Subsequently, we performed cavity ring-down spectroscopy on methane using the controllable comb source to demonstrate the applicability of the comb in spectroscopic analysis. A typical ring-down cavity comprises two high-reflectivity mirrors, between which an optical beam is stored. With each reflection, a small fraction of the stored light leaks out, resulting in an exponential decay of the intracavity light intensity, *I(t)*, over time *t*. Due to the exceptionally high finesse of the cavity, the effective light and matter interaction path length extends over several kilometers. The rate of this decay, known as the ring-down rate $\beta$, is governed by the reflectivity of the cavity mirrors and the presence of absorbing species within the cavity. The total ring-down rate is expressed as

$$\beta = \beta_0 + \sigma N c, \quad (1)$$
$$\beta_0 = c(1 - R)/L, \quad (2)$$

where $\sigma$ is the photo absorption cross-section of the intracavity absorbing species, $N$ is the number density, $c$ is the velocity of light in air, $\beta_0$ is the background ring-down rate without the intracavity absorbing species, $R$ is the reflectivity of the cavity mirrors, and $L$ is the cavity length. $N$ can be determined from the difference in the ring-down rates, $\beta - \beta_0$. Both $\beta$ and $\beta_0$ are obtained by fitting the temporal decay of the ring-down signal to an exponential function. One key advantage of CRDS is its insensitivity to shot-to-shot laser intensity fluctuations, because the decay rate is essentially unaffected by laser intensity.

The system configuration is shown in Fig. 8. The comb source was fiber-coupled to an optical isolator (Thorlabs, IO-H-1550APC), which was used to prevent the comb from back-reflecting light from the cavity mirrors. An AOM (Aerodiode, RFAOM-T-80) was used to interrupt the light and generate ring-down decay. The light beam was then modified using a collimator (Thorlabs, PAF2-A4C) to match the $TEM_{00}$ modes of the optical cavity. The cavity had a length of approximately 241 mm, yielded an FSR of 622 MHz, and was made of invar alloy to ensure an extremely small expansion coefficient. M1 and M2 were two high-reflection, low-loss cavity mirrors (Layertec), with a reflectivity of 99.9987 % and a finesse of 244,000. M1 was installed in a piezoelectric actuator (Thorlabs, PA44M3KW), such that the cavity length was dithered to observe each comb mode. Finally, the light beam was introduced into an NIR photodetector (Thorlabs, DET10N2) after passing through a focusing lens (f = 20 mm). The output signal of the photodetector was introduced into a comparator, which generated a transistor–transistor logic signal when sufficient photons were stored in the cavity to switch off the AOM and trigger the digitizer (National Instruments, PXIe-5922). The optical cavity was installed in a temperature-controlled enclosure, and the temperature was controlled using a temperature controller system (Thorlabs, TC300). The fluctuation of the temperature was effectively eliminated, with a total fluctuation of ~0.04 °C h$^{-1}$ during the experiment.

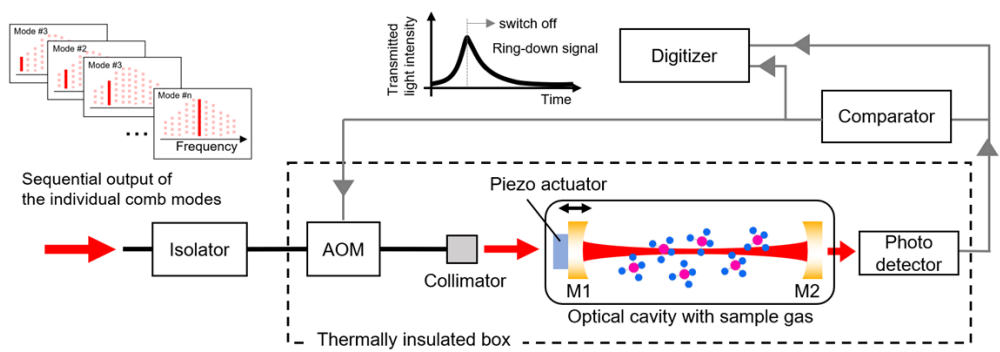

Fig. 8. Cavity ring-down system configuration with sequential output of the individual comb modes (M1 and M2: Cavity mirrors). A freely controllable frequency comb was scanned using a scanning time of 4 s for a spectral scanning range of 30 pm.

Unlike conventional comb-based CRDS, the ring-down signal of each comb tooth can easily be observed, similar to a typical narrow-bandwidth continuous-wave (CW) laser source. Figure 9a shows the ring-down decay measured at a fixed wavelength of 1650.96 nm in pure nitrogen. The vertical axis represents the measured signal voltage corresponding to the light intensity, and the horizontal axis represents the decay time. Least-squares exponential fitting was used to determine the ring-down time. The lower panel shows the residuals from the fit, with a standard deviation of $1.2 \times 10^{-3}$ V. The absence of structure indicates that the ring-down decay was well explained by an exponential function. The fitted ring-down time $\tau$ was 66.7 μs.

The Allan–Werle deviation of the ring-down rate was evaluated to assess system stability. For comparison, an external cavity diode laser (ECDL) operating at the same fixed wavelength and with an output power of 10 mW was employed. The signal acquisition rate $f$ was approximately 30 Hz. The Allan–Werle deviations are plotted together with the ideal Allan–

Werle deviations for white noise versus integral time, as shown in Fig. 9(b). The red and light blue lines correspond to the results with the freely controllable OFC and ECDL, respectively. The sensitivity of the CRDS was quantified using the minimum detectable absorption coefficient (MDAC) and noise equivalent absorption coefficient (NEAC) [44]. The MDAC obtained in a single shot is defined by $\alpha_{min} = \sigma(\tau)/(c\tau^2)$, and the NEAC is given by $\alpha_{min}(2f)^{1/2}$, where $\sigma(\tau)$ is the standard deviation of $\tau$, and $f$ is the repetition rate of the $\tau$ measurements.

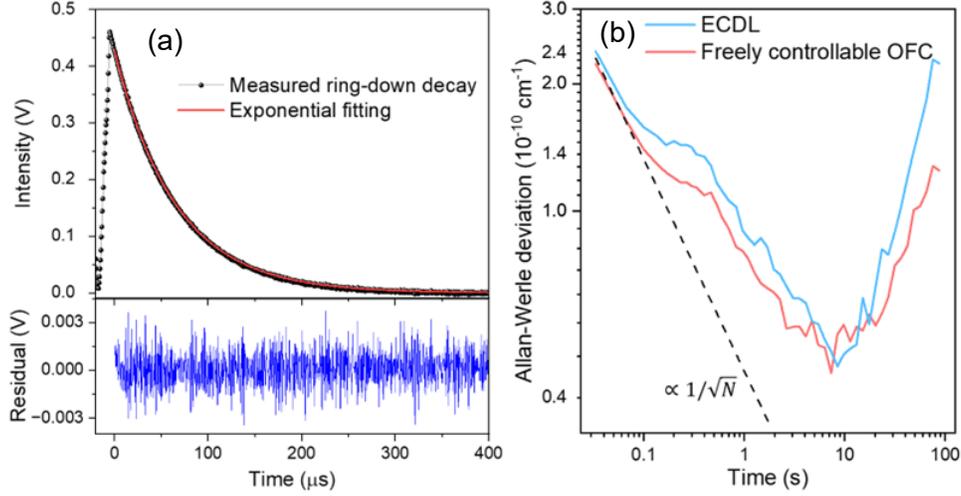

Fig. 9. (a) Upper panel: Measured ring-down decay using the frequency comb and exponential fitting, Lower panel: Residuals of the exponential fitting, (b) Allan–Werle deviations of the measured $(c\tau)^{-1}$ with freely controllable OFC (red line) and ECDL (blue line), and white noise for comparison (black broken line). $N$ is the sample number of the ring-down signal.

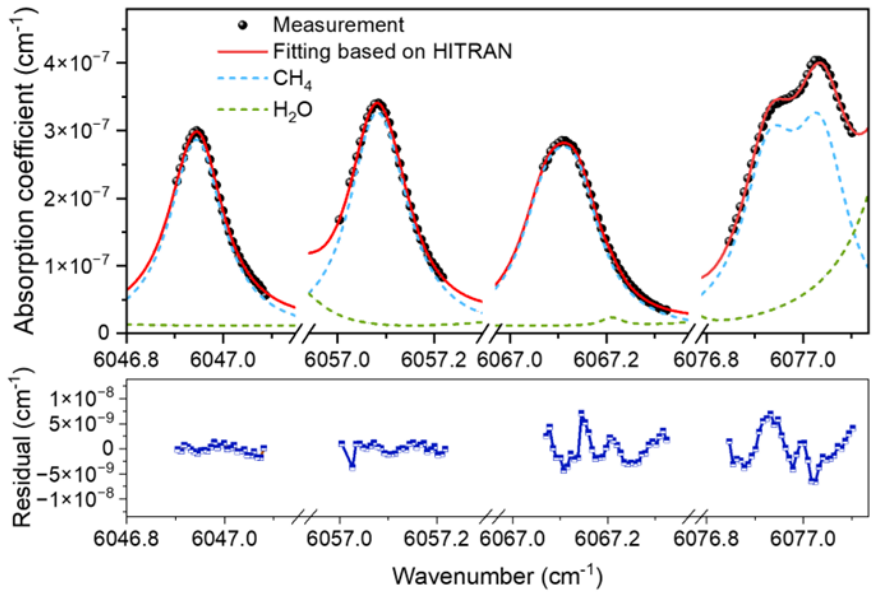

Fig. 10. Measured $CH_4$ absorption spectra at $2\nu_3$ bands with a concentration of 710 nmol/mol, least-squares fitting results (red line), and spectra components for CH4 (light blue broken line) and $H_2O$ (green broken line) calculated from the HITRAN database. The residuals (lower panel bule line) of $8 \times 10^{-10}$ cm$^{-1}$, $1 \times 10^{-9}$ cm$^{-1}$, $3 \times 10^{-9}$ cm$^{-1}$, and $4 \times 10^{-9}$ cm$^{-1}$ were obtained for the wavenumbers of 6046.9 cm$^{-1}$, 6057.1 cm$^{-1}$, 6067.1 cm$^{-1}$, and 6077 cm$^{-1}$, respectively.

In this setup, we obtained a single-shot MDAC of $2.2 \times 10^{-10}$ cm$^{-1}$ and an NEAC of $5.7 \times 10^{-11}$ cm$^{-1}$ Hz$^{-1/2}$ for freely controllable OFC. By averaging over 9 s, the MDAC was reduced to $4.2 \times 10^{-11}$ cm$^{-1}$, corresponding to a CH$_4$ sensitivity of 0.082 nmol/mol. Notably, the Allan deviation plot of a single comb tooth exhibited comparable performance to the ECDL, demonstrating that the implemented filtering scheme offers promising stability.

Methane absorption transitions at the $2\nu_3$ bands were then measured. The optical cavity was filled with 710 nmol/mol methane diluted with N$_2$ at atmospheric pressure, but only a small volume of H$_2$O was observed. The temperature of the cavity was precisely controlled to 24.5 ± 0.01 °C. To observe the absorption structure of CH$_4$, a freely controllable frequency comb was scanned by controlling the VIPA + DMD filter at a scanning time of 4 s over a spectral scanning range of 30 pm. The single-comb mode was selected individually with a resolution of 206 MHz, as mentioned previously. The wavelengths were tuned to approximately 1653.7 nm (6047 cm$^{-1}$), 1650.9 nm (6057 cm$^{-1}$), 1648.2 nm (6067 cm$^{-1}$), and 1645.5 nm (6077 cm$^{-1}$). Figure 10 shows the spectra obtained by averaging five scans (black dots), together with the fitting results from least-squares analysis using Voigt profiles (red line), and spectra for CH$_4$ (light blue broken line) and H$_2$O (green broken line) calculated using the HITRAN database [45] with Voigt profiles. Owing to the high resolution and high sensitivity, molecular absorption peaks were recorded with high accuracy. The residuals of the fitting are shown in the lower panel of Fig. 10, with standard deviations of $8 \times 10^{-10}$ cm$^{-1}$, $1 \times 10^{-9}$ cm$^{-1}$, $3 \times 10^{-9}$ cm$^{-1}$, and $4 \times 10^{-9}$ cm$^{-1}$ and the peak absorption coefficients of the 6046.9 cm$^{-1}$, 6057.1 cm$^{-1}$, 6067.1 cm$^{-1}$, and 6077 cm$^{-1}$ lines in 710 nmol/mol CH$_4$ respectively, demonstrating the high precision of the developed spectrometer for trace CH$_4$ analysis. Using a low-pressure gas sample, we accurately observed narrow and structured spectral shapes using the developed system.

## 3. Discussion

Table 1 presents a comprehensive comparison of the freely controllable OFC-CRDS system with previous DFC-CRDS implementations. The FTS-based DFC-CRDS [24] achieved a high-spectral resolution of 250 MHz. However, the acquisition of interferograms requires a long time, making them unsuitable for rapid detection. In contrast, mode-resolved DFC-CRDS enables the selective detection of individual comb lines, offering the potential for rapid measurements with multiple modes. We observed each comb line separately by tuning the PZT for an approximate FSR of 1 (622 MHz), which yielded the highest spectral resolution of ~206 MHz (corresponding to the $f_{rep}$ of the comb) reported for DFC-based CRDSs. Using a high-finesse cavity with an FSR close to $f_{rep}$ of the comb, it would be possible to acquire multiple ringdown signals more rapidly by slightly sweeping its cavity length. Furthermore, the arbitrarily controlled comb structure ensures a promising power distribution across individual comb lines and mitigates detector saturation issues that commonly arise in multimode

Table 1 Comparison of the reported direct frequency comb cavity ring-down spectroscopy (DFC-CRDS)

| OFC-CRDS references | Finesse | $f_{rep}$ | Spectral resolution | Sensitivity (averaging time) |
|---|---|---|---|---|
| Thorpe et al. [21] | 7200 | 100 MHz | 25 GHz | $2 \times 10^{-8}$ cm$^{-1}$ (1 s) |
| Lisak et al. [22] | 19800 | 1 GHz | 1 GHz | $2.6 \times 10^{-8}$ cm$^{-1}$ (1 s) |
| Chen et al. [23] | 20000 | 9.62 GHz | 9.62 GHz | $4.3 \times 10^{-8}$ cm$^{-1}$ (1 s)* |
| Dubroeucq et al. [24] | 21400 | 250 MHz | 250 MHz | $1.1 \times 10^{-9}$ cm$^{-1}$ (1 h 19 min) |
| This work | 244000 | 206 MHz | 206 MHz | $4.2 \times 10^{-11}$ cm$^{-1}$ (10 s) |

* This value was calculated by dividing the noise equivalent absorption by the square root of the number of comb modes ($M \approx 40$).

configurations [20,21]. This capability allows the implementation of an ultrahigh finesse cavity, considerably enhancing the measurement sensitivity. The system achieved an unprecedented sensitivity of $4.2 \times 10^{-11}$ cm$^{-1}$, two orders of magnitude better than previously reported comb-based CRDS systems.

## 4. Conclusion

In conclusion, we developed a freely controllable OFC source for highly sensitive spectroscopy based on the spectral peaking phenomenon. The comb modes that overlapped the CH$_4$ absorption spectra were transformed into background-suppressed spectral peaks in a nonlinear loop mirror equipped with a CH$_4$ gas cell. The spectral widths of generated spectral peaks were as narrow as ~32 pm at FWHM. Coherence-preserving power scaling of the generated comb was demonstrated using a fiber Raman amplifier. Only the single-comb mode was filtered using a newly developed ultrahigh-resolution spectral filter consisting of a VIPA and DMD. The maximum optical power of a single comb was estimated to be >10 mW after power scaling using an SOA. We demonstrated a freely controllable frequency-comb-based CRDS and achieved the highest spectral resolution for an DFC-based CRDS. The freely controllable frequency comb provided sufficient power-per-mode and ultranarrow linewidths, thus achieving a high-coupling efficiency, manifested by the increased build-up power in the high-finesse cavity with a short integration time. According to Allan–Werle deviation analysis, the DFC-CRDS system achieved a $4.2 \times 10^{-11}$ cm$^{-1}$ per spectral element following an integral period of 9 s, corresponding to a CH$_4$ sensitivity of ~0.14 nmol/mol. This sensitivity is two orders of magnitude higher than that of previously reported comb-based CRDS systems. The molecular spectrum can be observed at a high-spectral resolution limited by the $f_{rep}$ of the OFC. As a demonstration, we measured CH$_4$ absorption transitions at four different wavelengths. The filtered single-comb mode was scanned at 0.25 Hz and provided a spectral resolution of 206 MHz. The measured results showed that the system was capable of measuring CH$_4$ with extremely high precision, with ~0.29 % of fitting errors at a concentration of 710 nmol/mol.

This approach can be applicable to any wavelength range that can be generated by the optical comb, and is useful for monitoring other molecular gas species, such as NH$_3$, H$_2$O, CO, CO$_2$, and NOx, by changing the molecular gas cell or using a mixture of gases for the spectral peak generation. As the linewidth of a single comb is very narrow, a much higher sensitivity can be expected by using a higher finesse cavity. The comb mode can be controlled in a more functional manner; in this context, novel functional spectroscopic techniques can be realized in the near future.

## Data availability
Data underlying the results presented in this paper are not publicly available at this time but may be obtained from the authors upon reasonable request.

## Acknowledgments

This research was supported by JST Core Research for Evolutional Science and Technology (CREST) (grant number JPMJCR2104). The authors thank K. Hashiguchi, M. Amano, N. Ishiwata, M. Mukai, and K. Jung for the valuable discussions and cooperation.

## Conflict of interest

The authors declare no conflicts of interest.